\documentclass{aa}  

\usepackage{graphicx}
\usepackage{hyperref}

\usepackage{txfonts}
\usepackage{color}
\usepackage{caption}
\usepackage{multicol}

\begin{document}

   \title{IRAS4A1: Multi-wavelength continuum analysis of a very flared Class 0 disk}

   \author{O.~M. Guerra-Alvarado,
          \inst{1}
           N. van der Marel,
           \inst{1}
           J. Di Francesco,
           \inst{2}
           L.~W. Looney, 
           \inst{3}
           J.~J. Tobin,
           \inst{4}
           E.~G. Cox,
           \inst{5,6}
           P.~D.~Sheehan,
           \inst{5}
           D.~J. Wilner,
           \inst{7}
         E. Mac\'{i}as,
          \inst{8}
          C. Carrasco-Gonz\'{a}lez,
          \inst{9}
          }

   \institute{Leiden Observatory, Leiden University, PO Box 9513, 2300 RA Leiden, The Netherlands
        \\      \email{guerra@strw.leidenuniv.nl}
    \and      National Research Council of Canada, Herzberg Astronomy and Astrophysics Research Centre, 5071 West Saanich Road, Victoria, BC V9E 2E7, Canada
    \and          Department of Astronomy, University of Illinois, 1002 West Green Street, Urbana, IL 61801, USA
    \and          National Radio Astronomy Observatory, 520 Edgemont Road, Charlottesville, VA, USA
    \and          Center for Interdisciplinary Exploration and Research in Astronomy, Northwestern University, 1800 Sherman Rd., Evanston, IL, 60202, USA
    \and          NSF MPS-Fellow
    \and         Center for Astrophysics | Harvard \& Smithsonian, Cambridge, MA 02138, USA
    \and          ESO Garching, Karl-Schwarzschild-Str. 2, 85748, Garching bei Munchen, Germany
    \and           Instituto de Radioastronom\'{i}a y Astrof\'{i}sica (IRyA), Universidad Nacional Aut\'{o}noma de M\'{e}xico (UNAM)
}


 
  \abstract
   {Understanding the formation of substructures in protoplanetary disks is vital for gaining insights into dust growth and the process of planet formation. Studying these substructures in highly embedded Class 0 objects using the Atacama Large Millimeter/submillimeter Array (ALMA), however, poses significant challenges. Nonetheless, it is imperative to do so to unravel the mechanisms and timing behind the formation of these substructures.}
   {In this study, we present high-resolution ALMA data at Bands 6 and 4 of the NGC1333 IRAS4A Class 0 protobinary system. This system consists of two components, A1 and A2, separated by 1.8" and located in the Perseus molecular cloud at $\sim$293 pc distance.}
   {To gain a comprehensive understanding of the dust properties and formation of substructures in the early stages, we conducted a multi-wavelength analysis of IRAS4A1. Additionally, we sought to address whether the lack of observed substructures in very young disks, could be attributed to factors such as high degrees of  disk flaring and large scale heights. To explore this phenomenon, we employed radiative transfer models using RADMC-3D. We employed different approaches and compared the model outcomes with our observational data. This comparison allowed us to gain insights into the challenges in detecting substructures in nascent disks and shed light on the potential influence of the dust scale height on observations of protoplanetary disks.}
   {The continuum data revealed the presence of two disks/envelopes around A1 and A2, along with structure connecting the two sources. Furthermore, spectral index measurements indicate lower optical depth within the A2 disk compared to A1. Our multi-wavelength analysis of A1 discovered characteristics such as high dust surface density, substantial dust mass within the disk, and elevated dust temperatures. These findings suggest the presence of large dust grains compared to the ones in the interstellar medium (ISM), greater than 100 microns in size within the region. By employing RADMC-3D, we confirmed that increasing the scale height creates the appearance of an asymmetry in protoplanetary disks. Our findings indicate that a scale height of at least 0.3 (H/R) is necessary to produce this observed asymmetry. Furthermore,  while there's no direct detection of any substructure, our models indicate that some, such as a small gap, must be present. However, reproducing the intensity profile along the major and minor axes necessitates considering other processes that may be occurring within the IRAS4A1 disk.} {In summary, this result implies that disk substructures may be masked or obscured by a large scale height in combination with a high degree of flaring in Class 0 disks.}

\titlerunning{IRAS4A1: Multi-wavelength continuum analysis of a very flared Class 0 disk.}
\authorrunning{O. Guerra-Alvarado et. al.}
   \maketitle
%

\section{Introduction}
\begin{table*}[ht!]
 \centering
\Large
    \caption{Observations characteristics of IRAS4A}
 \setlength\tabcolsep{3pt}

\scalebox{0.65}{

\begin{tabular}{ccccccccccc}
 \hline
 \hline

  Project&             &ALMA&Repr.& ToS& & Sensitivity&Rms&Min BL&Max BL&BW\\
 Code & P.I  & Band&Frequency (GHz)&(s)& Array&(mJy)&(mJy)&(m)&(m)&(GHz)\\
 \hline

\multicolumn{11}{c}{ Long baseline observations}\\  
 \hline
  2018.1.00510.S& James Di Francesco & 6 &265.88  & 7111 & TM1(C43-8)&0.0220&0.10&92.1&8547.6&248-268 \\
   2018.1.00510.S& James Di Francesco & 4 &140.84  & 8913 & TM1(C43-9)&0.017&0.013&83.1&16196.3&136-154\\

 \hline
   \multicolumn{11}{c}{Short baseline observations}\\  
 \hline
  2018.1.00510.S& James Di Francesco & 6 &265.88  & 3267 & TM2 (C43-5)&0.04690&0.02050&15.1&2617.4&248-268 \\
   2018.1.00510.S& James Di Francesco & 4 &140.84  & 3125 & TM2 (C43-6)&0.037&0.045&15&2516.9&136-154\\

 \vspace{-0.3cm}

 \end{tabular}

  \label{tab:IRAS4A Observations}}
  \end{table*}
  
\begin{table*}[ht]
 \centering
 \Large
\caption{IRAS4A intrinsic continuum images characteristics}
\scalebox{0.65}{
\begin{tabular}{cccccccccc}
 \hline
  \hline
 				& Central& Central& \multicolumn{1}{c}{Synthesized beam}&&&&&& \\
& Frequency& Wavelength& Major $\times$ minor& Beam P.A. &rms&Peak flux A1&Peak flux A2&Robust&Peak SNR\\
 Band & (GHz)  & (mm)&(mas $\times$ mas)&(deg)&($\mu$Jy/beam)&(mJy/beam)&(mJy/beam)&&\\
 \hline


  6& 256.994 &1.2&78 $\times$ 31 & 20.82 & 135.299&14.04&19.92&0.0&147.5 \\
   4&  145.009 & 2.1&47$\times$29 & 5.79 & 26.27&3.13&2.63&0.0&120.4\\
    VLA Ka&  32.95 & 9.1 &75$\times$54& 79.94 & 9.773&0.719&0.306&0.5&80\\

 \vspace{-0.3cm}
 \end{tabular}
  \label{tab:HD163296 Observations especifications2}}
  \end{table*}
\begin{table*}[ht]
 \centering
 \Large
\caption{IRAS4A tapered continuum images characteristics}
\scalebox{0.65}{
\begin{tabular}{ccccc}
 \hline
  \hline
 				&   \multicolumn{1}{c}{Synthesized beam}&& \\
& Major $\times$ minor&rms&Peak flux A1&Peak flux A2\\
 Band&(mas $\times$ mas)&($\mu$Jy/beam)&(mJy/beam)&(mJy/beam)\\
 \hline


  6&78$\times$78 & 370.5 & 33.94&40.01 \\
   4& 78$\times$78 & 62.53 & 13.09&8.17\\
    VLA Ka& 78$\times$78& 9.89 & 0.943&0.355\\

 \vspace{-0.3cm}
 \end{tabular}
  \label{tab:HD163296 Observations especifications3}}
  \end{table*}
Recent studies of young stellar objects (YSOs) have concluded that dust growth from small particles to planetesimals may occur very early in the lifetime of protoplanetary disks (\citealt{2020A&A...640A..19T}; \citealt{2023ASPC..534..717D}). To stop the radial drift of dust particles and allow growth to happen, dust evolution models require dust particles to be trapped within disk substructures. This requirement implies that the formation of these substructures should be well underway during the Class 0/I phase of disk evolution. While such substructures have been detected in the disks of some Class I objects (\citealt{2020Natur.586..205S}; \citealt{2020Natur.586..228S}), these are still very limited in number. Moreover, they have not yet been observed in Class 0 protoplanetary disks. High-resolution studies of more evolved (Class II) protoplanetary disks, however, have revealed that substructures are common (\citealt{2015ApJ...808L...3A};\citealt{2018ApJ...869L..41A};\citealt{2016PhRvL.117y1101I}; \citealt{2018ApJ...869...17L}). These substructures, such as gaps, rings, arcs in cavities, and spiral arms (e.g., \citealt{2013atnf.prop.5561C}; \citealt{2013Sci...340.1199V}, \citealt{2016Sci...353.1519P}; \citealt{2018ApJ...869L..43H}), have proven crucial in reconciling the timescales for dust drift and planet formation, allowing larger dust particles to decelerate and grow further (\citealt{2012A&A...538A.114P}).

Many groups propose that disk substructures form due to interactions between the disk and forming planets (\citealt{2015ApJ...809...93D}; \citealt{2018ApJ...869L..47Z}). Other processes, however, may also contribute to their presence (\citealt{2015A&A...574A..68F}; \citealt{2015IAUGA..2256118Z}; \citealt{2016ApJ...821...82O}; \citealt{2018ApJ...865..102T}). The presence of substructures in evolved disks raises questions about when these substructures form and how are they linked to planet formation. The origin of these substructures, however, remains a subject of debate.

In addition to radial drift, vertical settling is another crucial process that significantly influences the evolution of dust particles in protoplanetary disks allowing them to grow and move into the mid-plane. This settling refers to the vertical motion of particles within the protoplanetary disk, driven by the balance between the gravitational force from the central star and the gas drag experienced by the particles. This settling process is influenced by several factors, including the turbulence of the disk and the sizes of the dust grains (\citealt{2004A&A...421.1075D}). In effect, larger dust grains are generally more efficient at settling due to their greater inertia, decoupling from the gas and moving into the mid-plane where they can grow further, while smaller grains experience stronger gas drag and tend to stay more mixed with the gas in higher layers of the disk (\citealt{2005A&A...443..185B}).
Vertical settling is expected to occur faster than radial drift and can be particularly pronounced in the inner regions of disks (\citealt{2014MNRAS.437.3055L}).
\begin{figure*}[t!]
\centering

\includegraphics[trim=0cm 4cm 0 0cm, clip=true,width=1.00\textwidth]{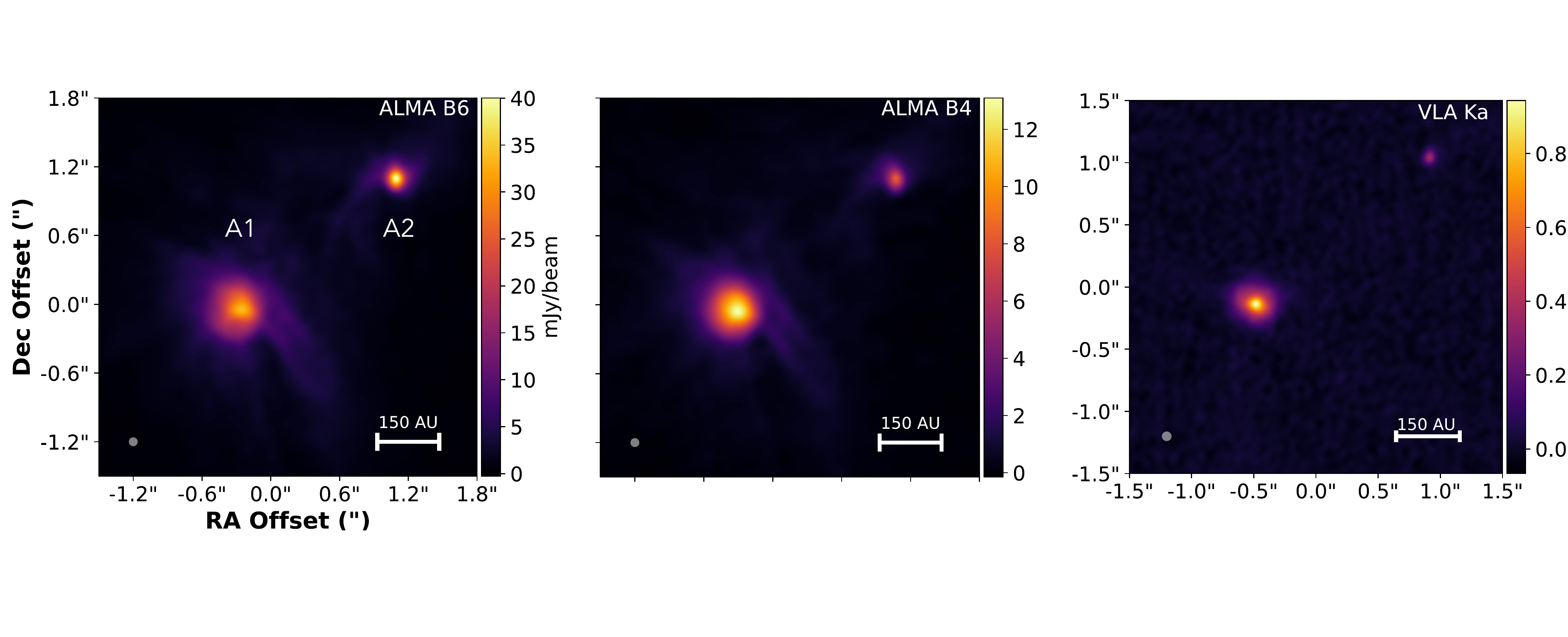}
\caption{Images of IRAS4A at 1.2 mm, 2.1 mm, and 9.1 mm imaged at 78 mas resolution. The central RA and Dec positions for Band 4 and Band 6 are 03:29:10.510 and +31.13.31.010, respectively. For the VLA image, the central Dec position is 03:29:10.502, and the central RA position remains the same, aligning with +31.13.31.010. Both sources are visible at a separation of 1.8" with some surrounded leftover emission seen between them at 1.2 and 2.1 mm but not at 9.1 mm (See appendix \ref{fig: Continuum Emission convolved at 78 m.a.s FE}, where the image color scale was changed to show the extended emission). Moreover, the peak emission of IRAS4A2 is larger than that of IRAS4A1 at 1.2 mm. The emission of both sources in the VLA image is more radially compact though it is very faint for IRAS4A2.} \label{fig: Continuum Emission convolved at 78 m.a.s}
\end{figure*}
In evolved Class II disks, some studies have observed that the larger dust particles are already primarily located at the mid-plane, indicating that settling has already occurred (\citealt{2016ApJ...816...25P}). On the other hand, the larger dust particles in younger disks may not have had enough time to settle completely, and the settling process may be ongoing in Class I disks with larger vertical extend (\citealt{2020A&A...642A.164V,2023ApJ...946...70V}).
Furthermore, previous studies have provided  enough evidence supporting the occurrence of grain growth in Class 0 young sources, as indicated by their low millimeter spectral indices. Notably, these studies have shown that the spectral index values are larger (ranging from 3.5 to 5) within the envelope at scales extending beyond 2000 au compared to the values (< 3.5) at smaller scales (<200 au) (\citealt{2009AAS...21341318K};\citealt{2007ApJ...659..479J}).

More recent findings suggest that young objects, particularly Class 0 YSOs, exhibit significant degrees of flaring and have considerable scale heights (\citealt{2022ApJ...934...95S}; \citealt{2022ApJ...937..104M}). This flaring and large scale height, irrespective of resolution or optical depth considerations, may conceal substructures in these systems or even prevent their formation. As a result, the settling of large particles is still ongoing, and much material remains in higher layers of the disk. More recently, a large study was performed in the eDisk survey \citep{2023ApJ...951....8O} of Class 0/I objects with several new findings about their young disks. 
Understanding the dust properties, vertical structures, and evolution of substructures from the early disk stages is crucial for comprehending the onset and progression of planet formation, as dust evolution and grain growth play vital roles in that process.

In this study, we investigate the protobinary system NGC1333-IRAS4A (IRAS4A), which is situated in the Perseus molecular cloud at a distance of 293 parsecs (pc) \citep{2018ApJ...869...83Z}. The system consists of two Class 0 protostars, namely IRAS4A1 and IRAS4A2, which are separated by an angular distance of 1.8" \citep{2018ApJ...867...43T}. Both IRAS4A1 and IRAS4A2 are surrounded by an envelope with a total mass of approximately 8 $M_{\odot}$ and a total luminosity of around 5 $L_{\odot}$ \citep{2019A&A...621A..76M}. Both objects have very well-distinguished outflows \citep{2015A&A...584A.126S}. We present here the continuum emission of both A1 and A2. We aim to study the structure of IRAS4A1 and investigate the absence of substructures in this component using radiative models. Additionally, in a separate paper, we will discuss the line emission and the presence of complex organic molecules in the IRAS4A system, as well as the continuum analysis of IRAS4A2.

\section{Observations}


%
\begin{figure*}[t!]
\centering

\includegraphics[width=\textwidth]{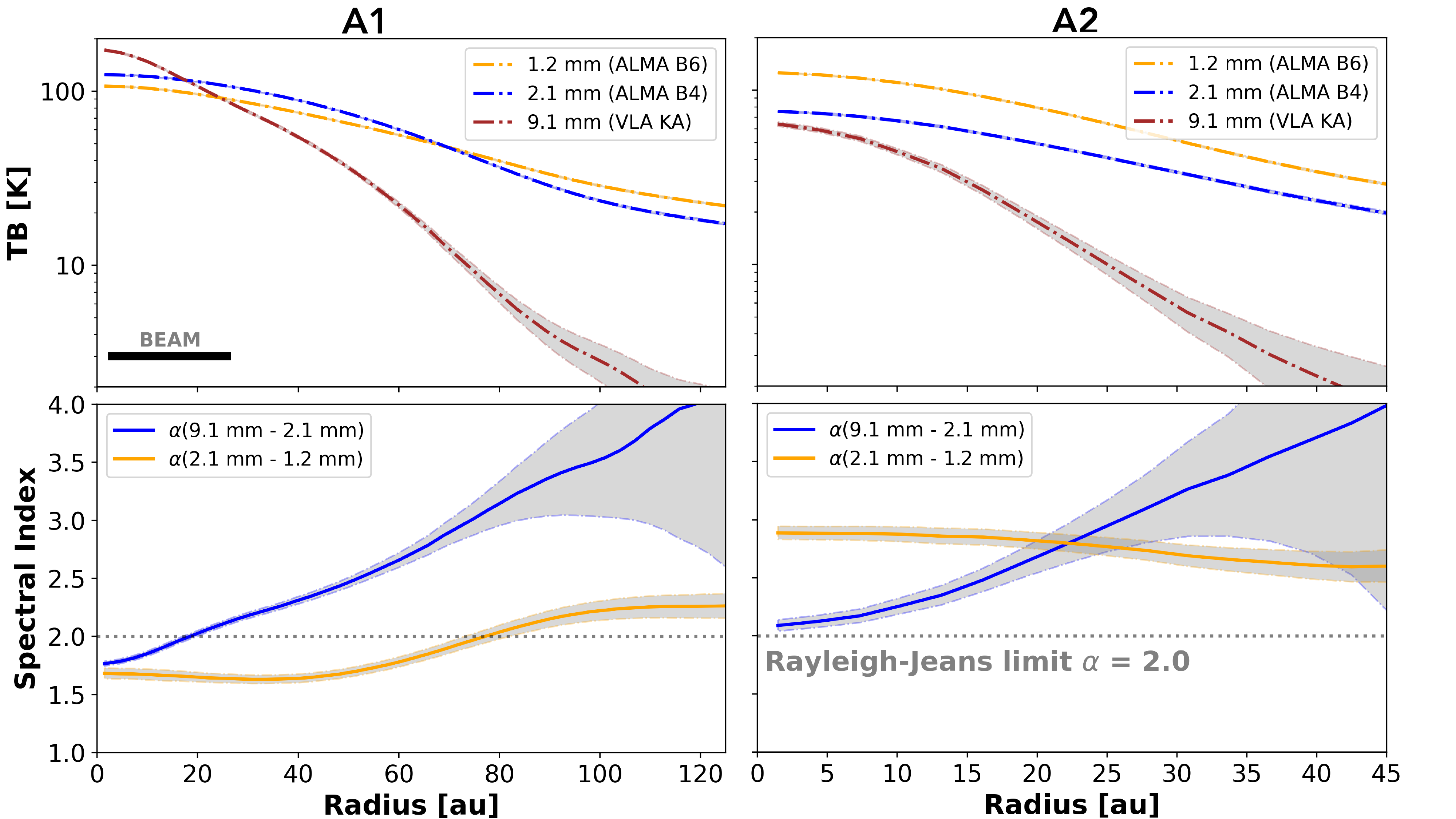}
\caption{ Upper right and left panels: Brightness temperatures at 1.2 mm, 2.1 mm, and 9.1 mm of A1 and A2, respectively. Lower right and left panels: Spectral indices between 9.1 mm - 2.1 mm and 2.1 mm - 1.2 mm of A1 and A2, respectively. In the inner parts of the A1 source, some free-free emission might be present at 9.1 mm. Moreover, the brightness temperature in A1 at 1.2 mm is lower than that at 2.1 mm which, as seen from the spectral index, might indicate very optically thick emission at those wavelengths and small dust particles (<1 mm). We only considered the statistical uncertainties for the brightness temperature and spectral index.} \label{fig: Bright temperatures and Spectral index of A1 and A2}
\end{figure*} 
The observations used in this paper were obtained using the Atacama Large Millimeter/submillimeter Array (ALMA). Band 4 (1.2 mm) and Band 6 (2.1 mm) data were taken as part of the project code 2018.1.00510.S (PI: James Di Francesco). The calibration of the data was performed by the ALMA staff and was restored by the allegro team at Leiden University. For Band 4, the observations were carried out in five execution blocks spanning from October 16th, 2018 to September 12th, 2021. Band 6 data were acquired in four execution blocks from November 19th, 2018 to September 30th, 2019. The total observing time on source for Band 4 was 3.34 h, while Band 6 had a total observing time of 2.88 h. Table 1 provides additional information regarding the characteristics of the data utilized in this study. 

The data reduction process was carried out using the Common Astronomy Software Applications (CASA, \citep{2007ASPC..376..127M}) version 5.7.0. The continuum spectral windows were separated from the line spectral windows and then averaged into eight channels for both data sets, Band 6 has 12 spectral windows centered at 264 Ghz, 252 GHz, and 250 Ghz with a total bandwidth of 2 GHz each. Band 4 has 15 spectral windows centered at 138 GHz, 150 Ghz, and 152 GHz with a total bandwidth of 2 GHz each. The flux calibration errors are set to the nominal values of 5$\%$ at Bands 4, 6.

Self-calibration techniques were employed for each spectral window individually using solution intervals of inf, 60s and 30s. Initially, we performed phase only self-calibration to the short baseline data  which resulted in significant improvements for Band 4 data (signal-to-noise ratio, from 88 to 380). For Band 6 data, however, only an increase in the SNR from 61 to 69 was achieved. In any case, sufficient self-calibration solutions were found, leading to enhanced data quality. Amplitude self-calibration was also performed but we stopped after a single step for most of the spectral windows, as it did not yield substantial improvements in the signal-to-noise ratio. For the long baseline data, we also performed phase self-calibration, although the improvement observed was comparatively less significant than in the short baseline data (Band 4 SNR from 42 to 86 and Band 6 SNR from 16 to 19). The reason for the lesser improvement in long baselines compared to short baselines could be attributed to a higher frequency of returns to the phase calibrator source during the long-baseline observations. Additionally, since self-calibration was exclusively applied in the same configuration, the cleaning process for long-baseline data often struggles to model the largest angular scales, even though they are present. This limitation affects the visibility data, especially considering the substantial amount of large scale emission present in these data. Amplitude self-calibration was only applied to a few specific spectral windows due to minimal enhancements in the SNR. The final data sets were obtained after concatenating all the spectral windows together in which no alignment was needed for any of them. 

Moreover, Very Large Array (VLA) data of IRAS4A was obtained from the VLA Nascent Disk and Multiplicity (VANDAM) survey \citep{2016ApJ...818...73T} conducted in the Perseus molecular cloud. The observations took place on October 21, 2013, employing the B-array configuration. For the correlator setup, two basebands with a total bandwidth of 4 GHz were utilized. These basebands were centered at frequencies of 36.9 GHz and 28.5 GHz, respectively. The setup was then further divided into 32 spectral windows each having a bandwidth of 128 MHz. The VLA Ka-band data in B-configuration has a shortest baseline length of 210 m and an estimated amplitude calibration uncertainty of $\sim$10$\%$. 

The final continuum images were created using task \textit{tclean} in CASA. In addition, we used the MTMFS deconvolver \citep{2011A&A...532A..71R} with nterms=2, together with scales of 0, 10, 30, and 50 times the pixel size (0.003" and 0.01" for ALMA and VLA images, respectively). Briggs weighting was found optimal for the purpose of this project, as it provided the best compromised between sensitivity and resolution, and several Robust values were explored when making the final images. Furthermore, for the Band 4 data, the uv range was modified to decrease the resolution. A smooth tapering was applied by setting \textit{uvtaper} to 0.058". Both, the Band 4 and Band 6 images were convolved to have the same 78 (milliarcsecond) beam. This common resolution allowed for a consistent analysis alongside the Very Large Array (VLA) data at 9.1 mm. Table 2 and Table 3 provide an overview of the characteristics of the images for Band 4 and Band 6, along with the VLA image obtained from the VANDAM survey. 

We acknowledge that the ALMA data for IRAS4A at Band 4, with its high resolution, time on source, and rms, can be favorably compared to the data obtained in the ALMA 2014 Long Baseline Campaign (LBC) Science Verification (SV) data of HL Tau at Band 6 (\citealt{2015ApJ...808L...3A}). The data for HL Tau was specifically designed to search for substructures, a goal that was also intended for the observation of IRAS4A1. The IRAS4A data set has a resolution of 47 mas, a time on source of 3.34 h and an rms of 13 $\mu$Jy, while the HL Tau data had a resolution of ~35 mas, a time on source of 4.5 h and an rms of 11 $\mu$Jy. Given the numerous substructures identified in the HL Tau disk and the comparable nature of the data, one would expect these observations to be sufficient for detecting substructures in the IRAS4A1 disk.

\section{Results}

Figure 1 displays the continuum images obtained from the observations. IRASA1 and IRAS2A2 are well resolved in both the Band 4 and Band 6 images. The majority of the submillimeter (sub-mm) emission originates from within each of these two sources. There is, however, additional faint emission observed between and surrounding A1 and A2, indicating some form of structure between the two sources. This structure is particularly evident at 2.1 mm and 1.2 mm wavelengths but not at 9.1 mm, which might be related to the lower sensitivity to thermal dust emission at 9.1 mm (see Appendix \ref{fig: Continuum Emission convolved at 78 m.a.s FE}). The origin of this extended emission remains unknown but the material could potentially be associated with the surrounding molecular cloud or with some diffuse envelope/core material at these scales. In contrast, the emission surrounding A1 or A2 is likely originating from the inner envelope or a very optically thick flared disk. Moreover, the brightness peak emission from A1 is lower than that from A2 at 1.2 mm, contrary to what is observed at 2.1 mm and 9.1 mm (see section 2 for the flux values). One possible explanation for this discrepancy could be that both sources have different scale heights and different optical depths. Despite the objects' similar age, Band 6 may be tracing different layers in A1 and A2, possibly not corresponding to the mid-plane. 

Furthermore, our ALMA images have been thoroughly examined, and no additional compact objects, such as low-mass companions or distant galaxies, have been detected within the field of view (> 3$\sigma$). Due to the high sensitivity and resolution of our data, it's highly improbable that any such objects have been missed. This suggests that A1 and A2 are unlikely to be part of a binary with a separation greater than 20 au, which is the long axis of the beam. However, is important to note that our data is not sensitive enough to detect a star lacking a circumstellar disk.

Lastly, a distinct asymmetry is observed for A1 in the 1.2 mm image, which is not apparent in the 2.1 mm and 9.1 mm images. The cause of this asymmetry warrants further investigation as it may provide valuable insights into the vertical structure of its Class 0 protoplanetary disk.

The radial profiles from these images were obtained by averaging the emission in elliptical rings for both sources and the central position of the radial profile of A2 was determined based on the peak emission in the respective images. Since there is an asymmetry in the A1 source, the central position of the radial profiles was determined by a Gaussian fit using \textit{imfit}. Although a slight bias might remain in the Gaussian fit, it was considerably less pronounced than using the peak emission center. Consequently, the center from the Gaussian fit is likely much closer to the ac- tual center of the source. For A1, the inclination and position angle values were set to 20 degrees and 96 degrees (NE direction, from North axis moving towards East), respectively, as reported for the outflow in \citealt{2016ApJ...819..159C}. On the other hand, for A2 we took the inclination to be 14 degrees \citep{2016ApJ...819..159C} while the position angle was taken from measurements on the inner outflow \citep{2021ApJ...916...82C} (122 degrees NE). Finally, the brightness temperature values were calculated by applying the full Planck equation to the radial intensity profiles as indicated by \citet{1979rpa..book.....R}, the concept refers to the temperature of a blackbody having the same brightness at that specific frequency. Moreover, two additional radial profiles were generated, representing the spectral indices between 9.1 mm and 2.1 mm, as well as between 2.1 mm and 1.2 mm. Figure 2 displays both the brightness temperature profiles and the spectral indices for A1 and A2.

By examining Figure 2, we can observe the behavior of the spectral indices for A1 and A2. For A1, the spectral indices are very low at the center of the source. As the radius increases, however, these indices gradually become larger. This continues until the noise of the 9.1 mm image starts to dominate the emission. Comparing the spectral index between 2.1 mm and 1.2 mm for A1 with that of A2, we find that the index for A1 remains consistently below 2 across most radii. On the other hand, the spectral indices for A2 are consistently above 2 throughout the range of radii considered. This discrepancy suggests that the emission from A1 is significantly more optically thick than that from A2 at these wavelengths. The difference in spectral indices between A1 and A2 implies differences in the physical properties of the two sources. For example, A1 may have a denser and more optically thick environment, which affects the observed spectral behavior. Furthermore, dust self-scattering might be affecting the inner regions of the IRAS4A1 disk. Additionally, some free-free emission might be increasing the brightness temperature in the inner regions of the A1 VLA image, affecting the spectral index between 9.1 mm - 2.1 mm.

In the recent study conducted by \citet{2019A&A...632A...5G} using independent measurements from the CALYPSO sample at 1.3 and 3.2 mm, they reported discovering remarkably low values of spectral indices (<2.0) within the inner regions of the IRAS4A1 envelope, specifically at distances of less than 200 au. This is in agreement with our spectral index values of the IRAS4A1 inner regions. Additionally, \citet{2019A&A...632A...5G} also observed higher spectral indices values extending up to 2000 au, which was attributed to grain growth processes occurring within the envelope. It is crucial to acknowledge that our high-resolution image  might be causing the extended component of the envelope to be resolved out, thereby making it difficult to measure the spectral index of this particular component. \citet{2007ApJ...659..479J}, previously pointed out that when extracting emission from the envelope, the spectral index of compact components would be flattened. Then,  spectral index values below 3.5 at smaller radii could be indicative of the presence of another component, most likely a disk.

\subsection{Multi-wavelength analysis of a Class 0 young stellar object.}
\begin{figure*}[ht!]
\centering

\includegraphics[width=600pt,trim={0cm 0cm 1cm 0cm},clip]{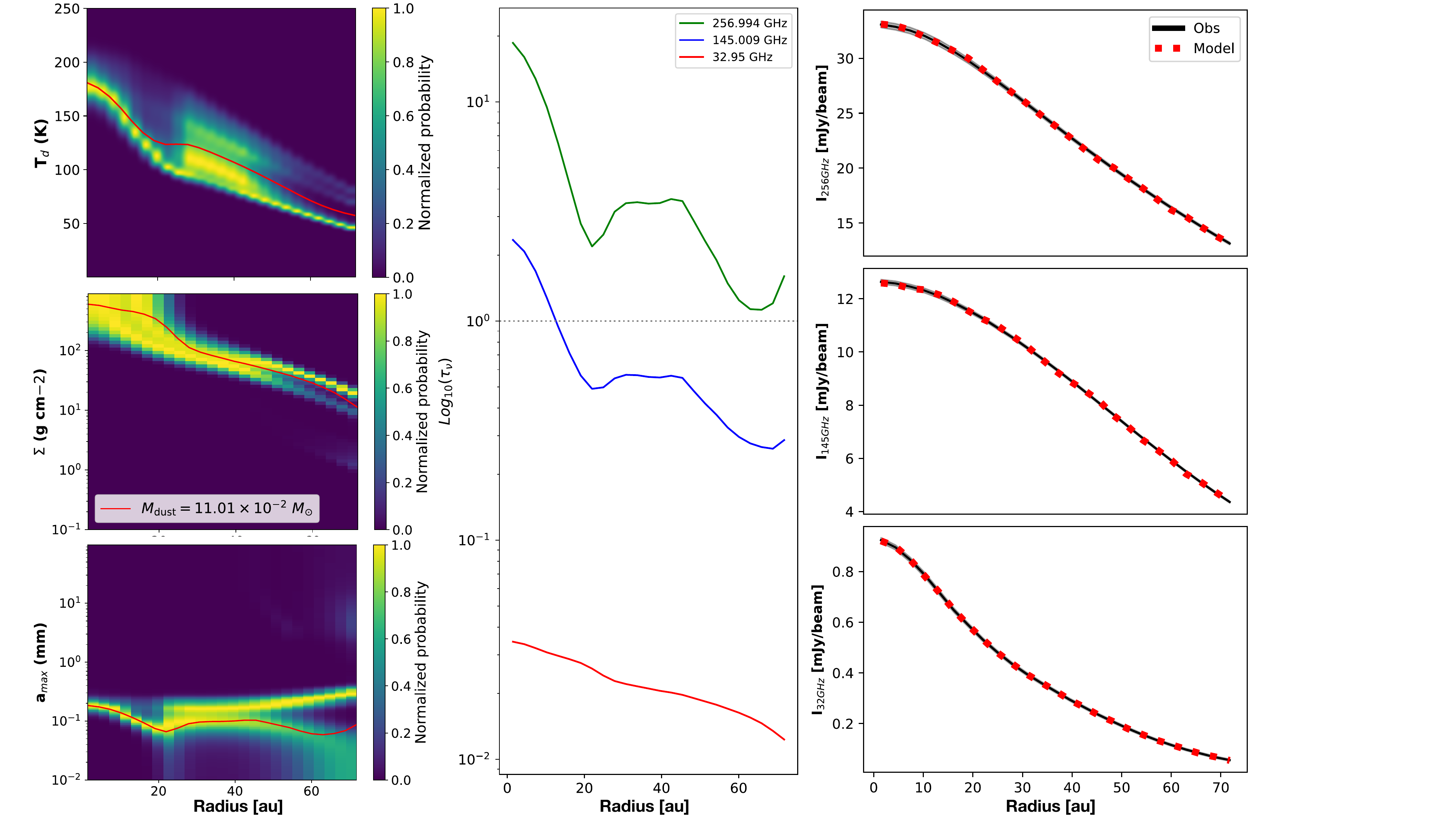}
\caption{ Left panels: Probability distributions of the dust parameters ($a_{max}$,$T_{dust}$,$\Sigma_{dust}$) of A1, the red lines in each plot is the expected value of each of the dust parameters. Middle panel: Optical depths at each radius of the three different wavelengths used in this work. Right panels: Comparison of the radial intensities from the observations and the model at each wavelength. The plots show very high temperatures and surface densities along the radius of the disk. Particle sizes have increased and are large in comparison with the ISM but still very small for the mid-plane of protoplanetary disks, in agreement with the expectation that not much settling has occurred in very young protoplanetary disks. Lastly, A1 shows optically thick emission at Band 6 and 4 at the inner radii.}  \label{fig: Probability distribution of the dust parameters}
\end{figure*}
In this study, we will adopt the hypothesis that the emission detected in our high-resolution images originates primarily from a disk rather than the envelope. The reason behind this is that the high resolution of our imaging may result in the loss of most of the emission from the extended components (envelope) and that, as mentioned before, our findings of low values of the spectral indices in Figure 2 further support the notion of a disk scenario. Of course, it needs to be noted that some emission coming from the inner envelope might still be contributing to the total emission. While there might still be some confusion within the envelope, we unfortunately didn't account for a dynamical distinction. Separating the continuum from the envelope is challenging, and due to the optical depth, analyzing the lines becomes quite limited. Additionally, to align with previous research, we will also consider the disk to be flared, similar to observations and results found in Class 0 Young Stellar Objects (YSOs) using ALMA data (\citealt{2022ApJ...934...95S};\citealt{2022ApJ...937..104M}) and as suggested by edge on observations (\citealt{2020A&A...642A.164V,2023ApJ...946...70V}) and recently, the eDisk survey \citep{2023ApJ...951....8O}.

Protoplanetary disks are commonly expected to have millimeter or even centimeter-sized dust particles. Because of such grain growth, the albedo of the dust can be high at millimeter wavelengths, indicating that scattering plays a significant role in the opacity of the dust emission. When scattering is a dominant factor, the spectral index of the dust emission can no longer be directly associated to a spectral index of the dust opacity (i.e., $\beta$) (e.g., \citealt{2020ApJ...892..136S}; \citealt{2019ApJ...877L..18Z}).To analyze the spectral energy distribution (SED) of protoplanetary disks properly, it is crucial to consider both absorption and scattering effects in the dust opacity. To include the scattering effect, we can write the source function in the radiative transfer equation as:
 \begin{eqnarray}
S_{\nu}(T)=\omega_{\nu}J_{\nu}+(1-\omega_{\nu})B_{\nu}(T),
  \label{eq: Source function with scattering}
 \end{eqnarray}
\noindent where $J_{\nu}$ is the local mean intensity and $\omega_{\nu}$ is the albedo, defined by the scattering coefficient and the absorption coefficient as $\omega_{\nu}=\frac{\sigma_{\nu}}{\kappa_{\nu}+\sigma_{\nu}}$. We can approximate this to the analytical solution found in \citealt{{1993Icar..106...20M}} assuming a disk as a vertically isothermal slab and with isotropic scattering:
 \begin{eqnarray}
J_{\nu}=B_{\nu}(T)[1 + f(t,\tau_{\nu},\omega_{\nu})],
  \label{eq:Approximation by Miyake & Nakagawa}
 \end{eqnarray}
\noindent where
 \begin{eqnarray}
f(t,\tau_{\nu},\omega_{\nu})=\frac{\exp (-\sqrt3\epsilon_{\nu}t )+\exp (\sqrt3\epsilon_{\nu}(t-\tau_{\nu})) }{\exp(-\sqrt3\epsilon_{\nu}\tau_{\nu})(\epsilon_{\nu}-1)-(\epsilon_{\nu}+1)},
  \label{eq:Approximation by Miyake & Nakagawa }
 \end{eqnarray}
\noindent where $t$ is the optical depth variable and $\tau_{\nu}$ = $\Sigma_{dust}\chi_{\nu}$, where both are measured perpendicular to the disk mid-plane. Also, $\epsilon_{\nu}=\sqrt{1-\omega_{\nu}}$. Considering inclination effects by correcting the optical depth by the inclination angle ($i$) of the disk, we reach the emergent specific intensity obtained by \citealt{2019ApJ...876....7S}:
 \begin{eqnarray}
I_{\nu}=B_{\nu}(T)[(1-\exp( \tau_{\nu}/\mu))+\omega_{\nu}F(\tau_{\nu},\omega_{\nu})],
  \label{eq:emergent intensity (Sierra et al. 2019)}
 \end{eqnarray}
where, 
 \begin{eqnarray}
F(\tau_{\nu},\omega_{\nu})=\frac{1}{\exp(-\sqrt3\epsilon_{\nu}\tau_{\nu})(\epsilon_{\nu}-1)-(\epsilon_{\nu}+1)}\textsc{x}\nonumber\\ \left [  \frac{1-\exp(-(\sqrt{3}\epsilon_{\nu}+1/\mu)\tau_{\nu})}{\sqrt{3}\epsilon_{\nu}\mu+1}+\frac{\exp(-\tau_{\nu}/\mu)-\exp(-\sqrt{3}\epsilon_{\nu}\tau_{\nu})}{\sqrt{3}\epsilon_{\nu}\mu-1}  \right ],
  \label{eq:F(tau,omega) (Sierra et al. 2019)}
 \end{eqnarray}

It is important to mention that for these equations isotropic scattering is assumed, which may be an incorrect approximation for $2\pi a\geqslant \lambda$. To reduce the effect of the approximation, we replace the scattering coefficient in all equations with an effective scattering coefficient in the form (\citealt{1978JOSA...68.1368I}, \citealt{2018ApJ...869L..45B}):
\begin{eqnarray}
\sigma^{eff}_{\nu}=(1-g_{\nu})\sigma_{\nu},
  \label{eq:Effective scattering coefficient}
 \end{eqnarray}
\noindent where $g_{\nu}$ is the asymmetry parameter, i.e., the expectation value of cos $\theta$, where $\theta$ is the scattering angle (e.g., \citealt{1978JOSA...68.1368I}, \citealt{2018ApJ...869L..45B}). The values of $g_{\nu}$ depend on the dielectric properties of the dust particles. For our calculations, the values obtained in \citealt{2018ApJ...869L..45B} for $\sigma^{eff}_{\nu}$ were used.

In our analysis, the particle size distribution is assumed to follow a power law with a slope ($n(a)\propto a^{-p}$), where p is commonly assumed to be 3.5 according to measurements of the ISM (\citealt{1977ApJ...217..425M}). Also, the DSHARP opacity data (\citealt{2018ApJ...869L..45B}) was employed, which considers particles without porosity and a composition of 20 $\%$ water fraction by mass, 32.91 $\%$ astronomical silicates, 7.43 $\%$ troilite, and 39.66 $\%$ refractory organics.  

Equation 5 then ultimately depends on only three free parameters: dust temperature ($T_{dust}$), the surface density ($\Sigma_{dust}$), and the particle size ($a_{max}$). With three or more observed wavelengths, it becomes possible to solve the equation and obtain estimates for the  three free parameters ($T_{dust}$, $\Sigma_{dust}$, $a_{max}$).

It is important to note that this model assumes a single temperature at each radius within the disk. This assumption generally holds when most of the dust is settled in the disk's mid-plane. In cases where the emission is originating from an envelope or a flared disk involving different layers, however, this assumption may not be valid. So, it is worth noting that the temperature structure within protoplanetary disks can be complex, particularly if there are significant vertical temperature gradients or if different layers of the disk are contributing to the observed emission at different wavelengths. In these situations, a more sophisticated modeling approach that considers the vertical structure and temperature gradients within the disk would be necessary to interpret the observed SED. 

A multi-wavelength analysis similar to ours here was previously performed before on HL Tau using four images between 8 mm and 0.9 mm \citep{2019ApJ...883...71C} by simplifying the spectral behavior of the extinction coefficient using a power law. After that, it has been used in several other papers (e.g. \citet{2021A&A...648A..33M}, \citet{2021ApJS..257...14S} and \citet{2022A&A...664A.137G}) using the exact values of the dust opacity at each wavelength, including the work presented in this paper as well. This model is a first approach in determining the dust properties around a Class 0 YSO like A1.\\

A Bayesian approach was employed to obtain the posterior probability distributions of the model parameters ($a_{max}$,$T_{dust}$,$\Sigma_{dust}$) at each radius. To achieve this, a standard log-normal likelihood function was used, which is defined as follows:
\begin{eqnarray}
\ln p(\bar{I}(r)\mid \Theta )=-0.5 \sum_{i}^{}\left ( (\frac{\bar{I_{i}}-I_{m,i}}{\hat{\sigma}_{\bar{I,i}}})^{2}+\ln(2\pi\hat{\sigma}^{2}_{\bar{I,i}})  \right ),
  \label{eq:Log-normal likelihood function}
 \end{eqnarray}

\noindent where $\bar{I}$ is the azimuthally averaged intensity at radius $r$ and at frequency $\nu_{i}$, $I_{m,i}$ is the model intensity from different combinations of the three free parameters at a radius $r$, $\Theta$ is the vector of the three free parameters. In addition, we assumed that the uncertainty $\hat{\sigma}_{\bar{I,i}}$ at radius $r$ is:
\begin{eqnarray}
\hat{\sigma}_{\bar{I,i}}=\sqrt{\sigma^{2}_{\bar{I,i}}+(\delta_{i}\bar{I_{i}})^{2}},
  \label{eq:Uncertainty}
 \end{eqnarray}

\noindent where $\sigma^{2}_{\bar{I,i}}$ is the error of the mean, obtained from the azimuthally averaged intensity profiles (See section 2), and $\delta_{i}$ is the flux calibration error at each frequency. 

Figure 3 shows the analysis we performed, a model grid of intensities was created using various dust parameters. To infer the physical parameters of the dust particles, we compared the observed intensity at each radius with the expected spectral energy distribution (SED) derived from different combinations of the three free parameters in equations 5 and 6 ($a_{max}$ from 0.001 - 10 cm, $T_{dust}$ from 0.1 - 250 K, $\Sigma_{dust}$ from 0.1 - 1000 $gcm^{-2}$). In order to better match the observational data, the probability distribution of each parameter was plotted, along with the corresponding expected value (represented by the red curve in Figure 3). The expected value of each parameter was obtained by:
\begin{eqnarray}
E(X)=\frac{\sum_{i} X_{i}P(X_{i}|\bar{I}(r))}{\sum_{i}P(X_{i}|\bar{I}(r))}
  \label{eq:Expected value}
 \end{eqnarray}
\noindent where $X_{i}$ is each value in all the parameters inside our grid, and $P(X_{i}|\bar{I}(r))$ is the marginalized posterior probability of each parameter in every single cell of the grid. 

Multiple equally likely solutions close to each other were found for the A1 source, which equally explained the observed intensity. All the possible solutions have similar $\Sigma_{dust}$ and dust temperature, which is explained in \citet{2023ApJ...953...96Z} as there are no strong Mie interference patterns when $2\pi a<\lambda$. Finally, the optical depth values were derived from the analysis to provide insights into the dust properties at different locations within the disk. Figure 3 shows the dust parameters, the optical depths at each wavelength, and the intensities of the observations compared with the ones obtained from the model.

From Figure 3, it is evident that the A1 disk exhibits high optical thickness at the inner radii of A1, which poses challenges in fitting the dust parameters accurately. This observation suggests that the disk is highly unstable and contains very small dust particles (hundreds of microns in size) relative to dust grain sizes in protoplanetary disks. Notably, the derived temperature from the multi-wavelength analysis in A1 appears to be higher compared to other Class II disks analyzed using similar methods (\citet{2021A&A...648A..33M}, \citet{2021ApJS..257...14S}, \citet{2019asrc.confE..88G} and \citet{2019ApJ...883...71C}). This discrepancy may be attributed to the young age of the source and other processes occurring within the system, like infalling material that can contribute to the elevated temperature of the dust particles, viscous heating or even back warming by the envelope (\citealt{1993ApJ...412..761N}).

Furthermore, A1 displays a notably high dust surface density and mass in comparison to Class II disks. This result aligns with the notion that a significant portion of the material remains distributed as sub-mm particles surrounding the star rather than having settled and grown in the disk's mid-plane where it cannot be detected by our observations due to high optical depths, the high mass inferred is expected for a very young source like A1, which is likely to be very gravitationally unstable having still a substantial circumprotostellar mass not yet accreted by the central star. We note that the particle sizes found in other disk studies often vary significantly depending on the presence of substructures, which are not detectable in the A1 source. Moreover, the particle sizes observed in other Class II disks tend to be larger (cm-sized particles) compared to the 0.1 mm particles found in A1. This disparity can be attributed to the different evolutionary stages of the disks, with the dust in the other disks having evolved and settled more in the mid-plane.

Figure 3 indicates that the material flowing in to form the disk already contains large dust particles (> 10 microns) compared to the average ISM dust sizes. This suggests widespread grain growth across the entire disk radius. However, as one approaches the midplane and the central star, particles tend to become larger. These large dust particles compared to the ISM particles imply that grain growth is not limited to the midplane but also occurs in the flared regions of the disk where infall is the likely process that triggers this growth. Additionally, the increase in error at the outer region of the A1 disk is a result of the spatial sensitivity of the VLA image.\\

We compare the temperature profile of A1 with other Class 0 sources and found that A1's temperature agrees with those derived from CO and $H_{2}CO$ snowlines in IRAS04302 (Class I) and L1527 (Class I/O) by \citet{2020A&A...633A...7V}. 
Comparing with models from \citet{2017ApJ...835..259Y} on the Class 0 Protostar BHR71; however, we note that the derived densities in A1 are at least an order of magnitude lower. This difference could potentially stem from the observations utilized by \citet{2017ApJ...835..259Y} are of shorter wavelengths (\textit{Herschel}) that are more sensitive to the cloud, the surrounding envelope, and smaller dust grains.

Concerning particle sizes, our analysis indicates that the inner disk of A1 comprises particles nearly 0.3 mm in size. This suggests that the dust size distribution in the disk is primarily characterized by larger particles when compared to typical interstellar medium (ISM) dust sizes. However, in comparison to pebbles found in more evolved disks, these particles are relatively small. This suggests that although some dust growth has already occurred, the process is still ongoing. Several studies focused on Class 0 objects have measured dust sizes in the envelope using low dust emissivity indices, revealing that grain growth might already happen in this Class 0 objects maybe even up to mm-sized particles (\citet{2009A&A...507..861J},\citet{2019A&A...632A...5G}). More specifically, scattering measurements from polarization observations in IRAS4A show the possibility of large millimeter size particles within the system (\citealt{2015ApJ...814L..28C}). These findings diverge from the multi-wavelength analysis in our work that shows smaller dust particles in IRAS4A1.

The findings presented in Figure 3 do not provide a definitive explanation for the spectral index below 2 in Figure 2. The particle sizes around 0.1 mm align with what is expected for dust self-scattering, indicating the presence of low spectral indices \citep{2019ApJ...877L..22L}. Nevertheless, these observed values can also be rationalized by considering a highly optically thick disk within 
r<60 AU, where the inner layers are warmer than the outer layers. This scenario not only aligns with the observations but also corresponds to the outcomes illustrated in Figure 3, showcasing the high optical depth across all radii. 

\subsection{Generic gap models with large scale heights.}
\begin{figure*}[ht!]
\centering

\includegraphics[width=350pt,trim={0cm 3cm 0cm 0
cm},clip]{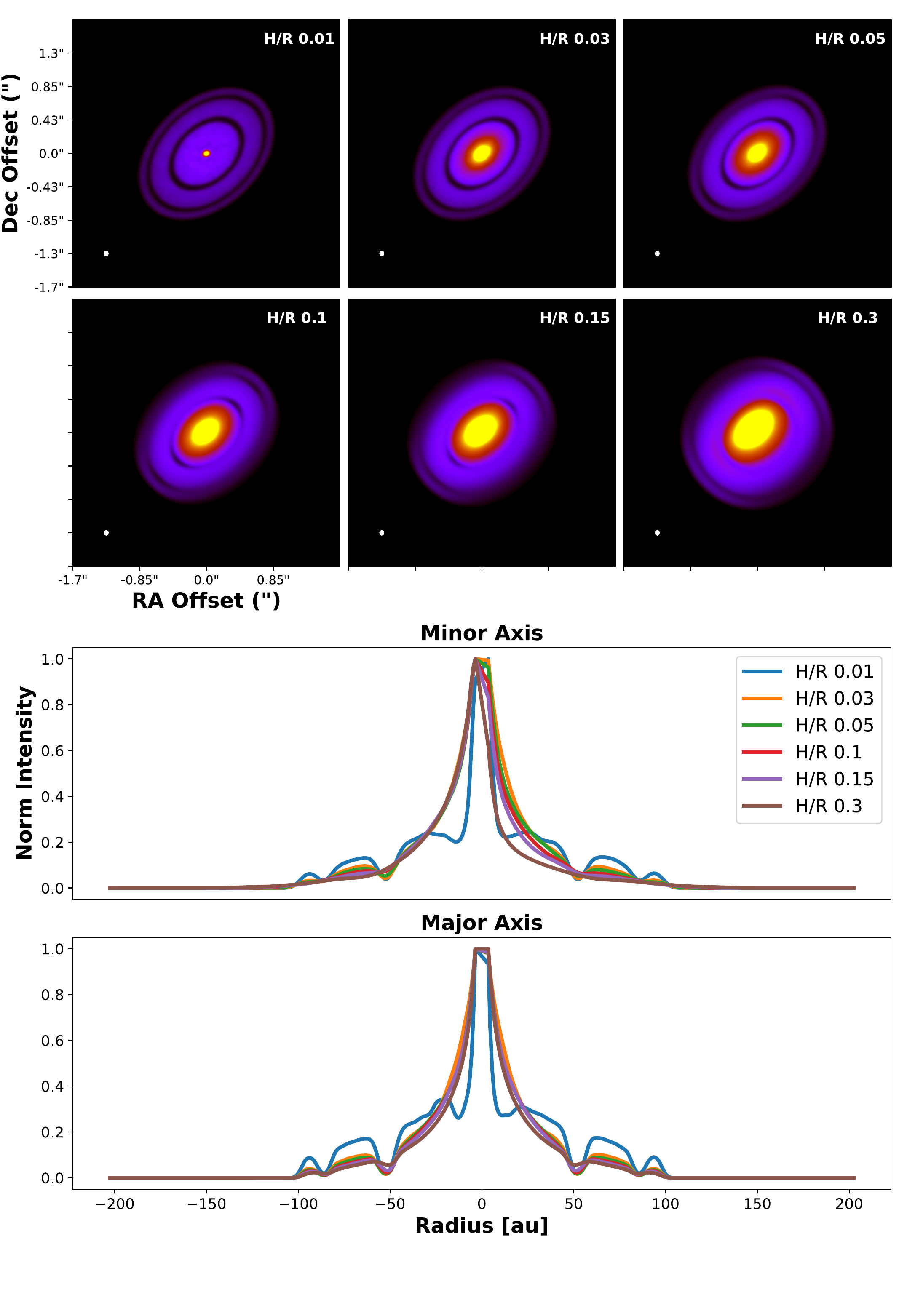}
\caption{ Top panels: Generic gap models in RADMC-3D increasing the scale height up to H/R 0.3. Bottom panels: Normalized intensities of the major axis and minor axis cuts of the generic gap models. An asymmetry in the minor axis of the disk becomes more prominent as the scale height increases. In both the minor and major axes, substructures start to flatten with increasing scale height, and after 0.3 they become barely visible.} \label{fig: IRAS4A1 RADMC-3D and major minor axis cuts}
\end{figure*}

\begin{figure*}[ht]
\centering

\includegraphics[width=\textwidth,trim={0.0cm 0.5cm 0cm 0
cm},clip]{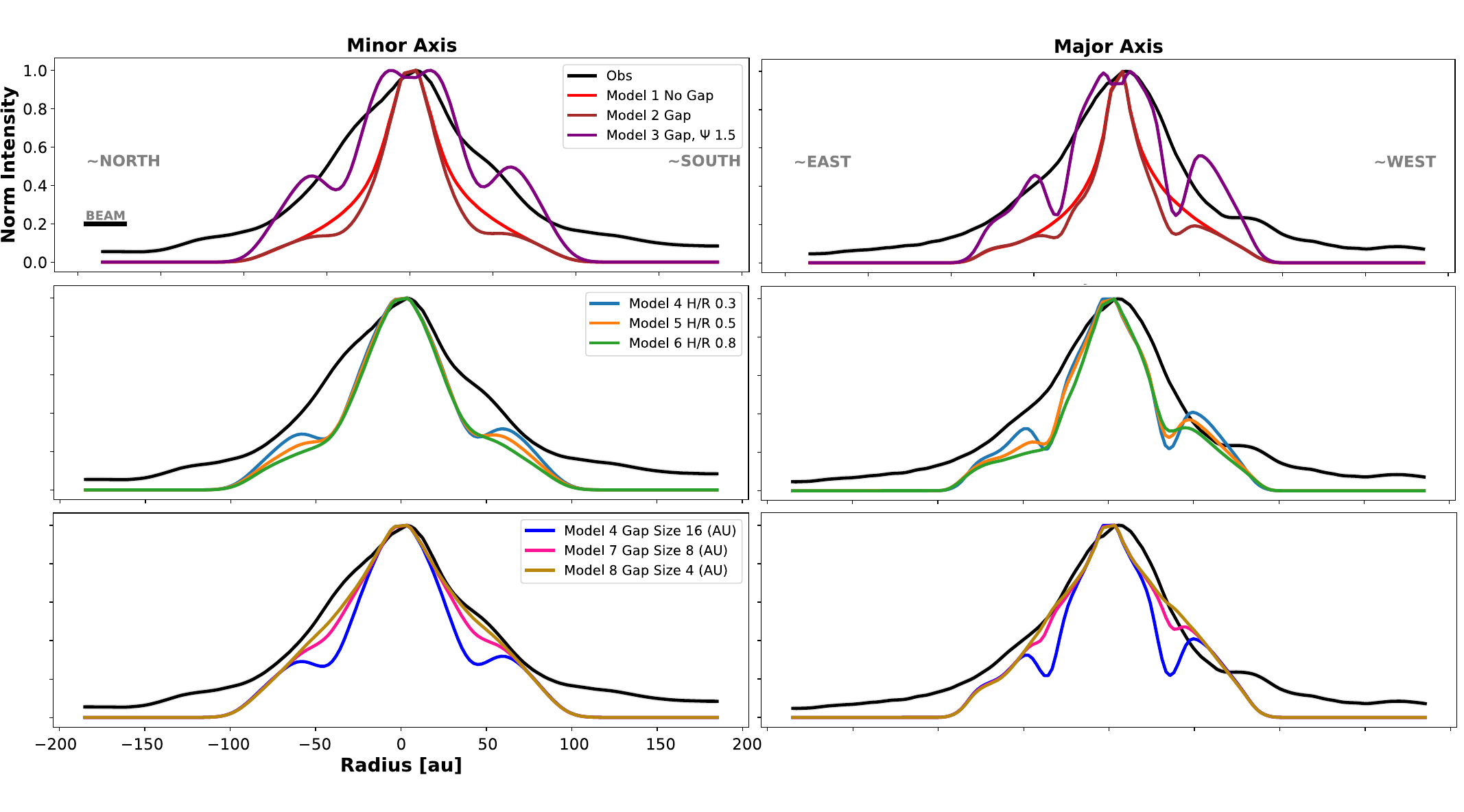}
\caption{Normalized intensities of the major axis and minor axis cuts of the IRAS4A1 RADMC-3D models. Top panels: The scale height of each model was fixed to H/R=0.3. Middle panels: The $\Psi$ flaring index values were fixed to 1.3 in the RADMC-3D models. Bottom panels: The scale height was fixed to 0.3 and the flaring index to 1.3, but the gap width was reduced in each model. The 'shoulder' shape and the asymmetry of the disk are better reproduced using a large scale height, a large flaring index, and a small gap width. Unfortunately, an excess of emission is found in the models and the small size of the gap needed cannot be observed at the current resolution.} \label{fig: IRAS4A1 RADMC-3D and major-minor axis cuts}
\end{figure*}
To explain the absence of observed substructures and the observed asymmetry in the IRAS4A1 disk, we employed radiative transfer models using RADMC-3D \citep{2012ascl.soft02015D}. For these models we assume that instead of observing a highly optically thick envelope with an embedded disk, IRAS4A1 is actually a flared disk with a significant scale height (the surrounding envelope has been resolved out in these high-res images, See section 3). When considering a greater scale height and flaring in the disk, it's crucial to differentiate between flared disks, which represent an equilibrium configuration of orbiting material, and an infalling model. In this study, we will model the flared disk solely from the perspective of the dust continuum, without incorporating a dynamic approach such as infalling or rotational motions. This flaring effect can create an asymmetry in the disk and make it challenging to detect substructures, if they exist. The combination of disk inclination, large scale heights, and optically thick emission, even at Band 6, contributes to this effect. Evidence supporting the presence of a highly flared disk instead of an envelope has been observed in the Class 0 Protostar L1527 IRS by \citet{2022ApJ...934...95S}.

To test this assumption, a model of the dust continuum emission at 1.2 mm was constructed using RADMC-3D. Initially, "generic gap models" inspired by the disk of HD163296 were made to investigate the disappearance of substructures with increasing scale height. Subsequently, we developed a specific model to the IRAS4A1 disk to reproduce the observed asymmetry at 1.2 mm in combination with the absence of substructures. These radiative transfer models allow us to perform a detailed examination of the disk's dust evolution and provide insights into its vertical structure.

For the generic gap models, we fixed certain parameters based on previous studies of HD163296. For the star, the parameters in Table 1 from \citep{2018ApJ...869L..41A} were used: $M_{\ast}$ = 2.04 $M_{\odot}$, $L_{\ast}$ = 17 $L_{\odot}$, $T_{\ast}$ = 9332 K, and a distance of 101 pc. The positions of the two most prominent gaps were taken to be 49 and 86 au with a fixed width of 10 and 8 au, respectively. 

The disk model was taken to have an inclination of i = 46.7°, a position angle of 133.3°, and a dust mass of 0.039$_{\odot}$ from \citet{2020A&A...633A.137D}. In addition, a size of 110 au for the disk was chosen. Inside RADMC-3D, a generic protoplanetary disk model was used, with the scale height varied in each model. To incorporate the DSHARP dust particle opacities, the optool software \citep{2021ascl.soft04010D} was utilized, allowing for their utilization within RADMC-3D. Finally for completeness, RADMC-3D calculated the dust temperature using the density distribution for the generic protoplanetary disk model as follows:

 \begin{eqnarray}
 \rho(r,z) =\frac{\sum(r)}{H_{p}\sqrt{2\pi}}exp(-\frac{z^{2}}{2H^{2}_{p}}),
 \label{eq: Density distribution}
 \end{eqnarray}
\noindent, where r is the distance to the star from the disk, $\sum(r)$, is the dust surface density, and $H_{p}$ is the scale height of the dust disk.\\

The scale height ($H_{p}$) in the generic protoplanetary disk model follows a power-law dependence on the radial distance as follows:

 \begin{eqnarray}
H_{p}=H_{100}(\frac{r}{100 AU})^{1+\Psi},
 \label{eq: Scale Height}
 \end{eqnarray}
\noindent where $\Psi$ is the flaring index, with a predefined value 0.14, and $H_{100}$, is the value of the scale height at a distance of 100 au from the central star.\\ 

The scale height parameter was increased in the generic gap models until the substructures disappeared due to shadowing, obscuration, and/or contrast effects. Figure 4 shows the images of these models together with a cut through their major and minor axes.

In Figure 4, it is evident that substructures present in young Class 0 disks are challenging to observe, if present, due to the large scale heights that these disks may exhibit. The cuts shown in Figure 4 provide additional insights into the behavior of the disk at different scale heights. Along the major axis, even at a low scale height of 0.05, a strong flattening effect on the rings and gaps is observed. This effect is highly dependent on the inclination and position angle of the disk. The intensity variations along the minor axis reveal another interesting aspect: in the SW part of the disk, a lack of intensity is observed, an asymmetry caused by a large vertical structure in the disk, also seen in other sources, such as \citet{2021ApJ...910...75L}, \citet{2023ApJ...951....9L}. The direction of this asymmetry is determined by the orientation of the modeled disk. Furthermore, as the scale height increases, both the depth of the gaps and the visibility of substructures begin to flatten along the minor axis too. Eventually, there is a point where substructures ($\leq$10 au) can no longer be distinguished. This example demonstrates the impact of a highly flared disk on the visibility and discernibility of substructures, if any, in a young protoplanetary disk.

\subsection{Large scale height and very flared  disk models of IRAS4A1}

To investigate the asymmetry observed in the 1.2 mm image of the IRAS4A1 disk, additional modeling was performed in RADMC-3D. The objective was to determine whether or not the observed asymmetry could be reproduced in a large scale-height flared disk scenario. To set up the RADMC-3D models, we fixed specific parameters. Due to the difficulty of determining the stellar properties directly from the literature for a highly embedded Class 0 object like IRAS4A1, average values of stellar properties in a number of Class I systems were obtained from Tables 1 and 2 in \citet{2023ApJ...944..135F}. These average values include the stellar mass (1.55 $M_{\odot}$), radius (2.1 $R_{\odot}$), and effective temperature (3700 K). The inclination and position angle of the disk were fixed at 20° and 99°, respectively. The dust mass in the disk was taken from the multi-wavelength analysis, resulting in a value of 0.11$^{+0.08}_{-0.04}$ $M_{\odot}$. The scale height in the RADMC-3D models for the IRAS4A1 disk was initially set to H/R = 0.3, based on the appearance of asymmetry in the generic gap models. In addition to this base model, eight more models were created. three with a fixed scale height, three with a fixed high flaring index ($\Psi$ = 1.3), and two models with reduced gap widths. This variety allowed exploring different scale heights within the context of a consistently high flaring profile. Figure 5 shows the corresponding cuts through the major and minor axes in all eight models. The IRAS4A1 observation and the model that best reproduce its intensity along the major and minor axis are shown in Figure 6.  By examining the outcomes of these various models, we can observe the influence of a gap presence, large flaring index, and large scale heights on the observed asymmetry and young Class 0 sources like the IRAS4A1 disk.

From the radiative transfer models of IRAS4A1, it is evident that an asymmetry is formed on the North (compared to South) part of the disk at large scale heights. The inclination and position angle in the models greatly influence the resulting asymmetry, emphasizing the uncertainties in these results. Furthermore, the simplicity of the model employed in this study may limit its ability to reproduce accurately the complexities of a Class 0 young stellar object like IRAS4A1. Nevertheless, the intensity profiles along the major and minor axes suggest the presence of "substructures" or other unknown processes occurring in the actual observations, as most models appear flat unless a gap is included. In the upcoming paragraphs, we will speculate about the substructure scenario in the IRAS4A1 disk although it is possible that something else is shaping the intensity profiles along the major and minor axis. 

The models with a very small gap exhibit intensity profiles that more closely resemble the observed profile at 1.2 mm in both the major and minor axes. This difference may indicate that the gaps at these early stages are still forming and that we will need still higher resolution to see them. Regardless of whether IRAS4A1 is indeed a flared disk, a lower limit on the scale height for generating an asymmetry can be established (H/R > 0.3). Note that the intensity profiles in the Figure are normalized, as the primary goal of this study is not to replicate the flux of the IRAS4A1 source precisely, but rather to provide insights into the earliest stages of disk and planet formation. Nevertheless, if our observations are capturing emission from higher layers in the disk and if the emission remains highly optically thick, it may be challenging to detect substructures with ALMA at the available resolution. 
\begin{figure*}[h!]
\centering

\includegraphics[width=1.05\textwidth,trim={1cm 3cm 0cm 0
cm},clip]{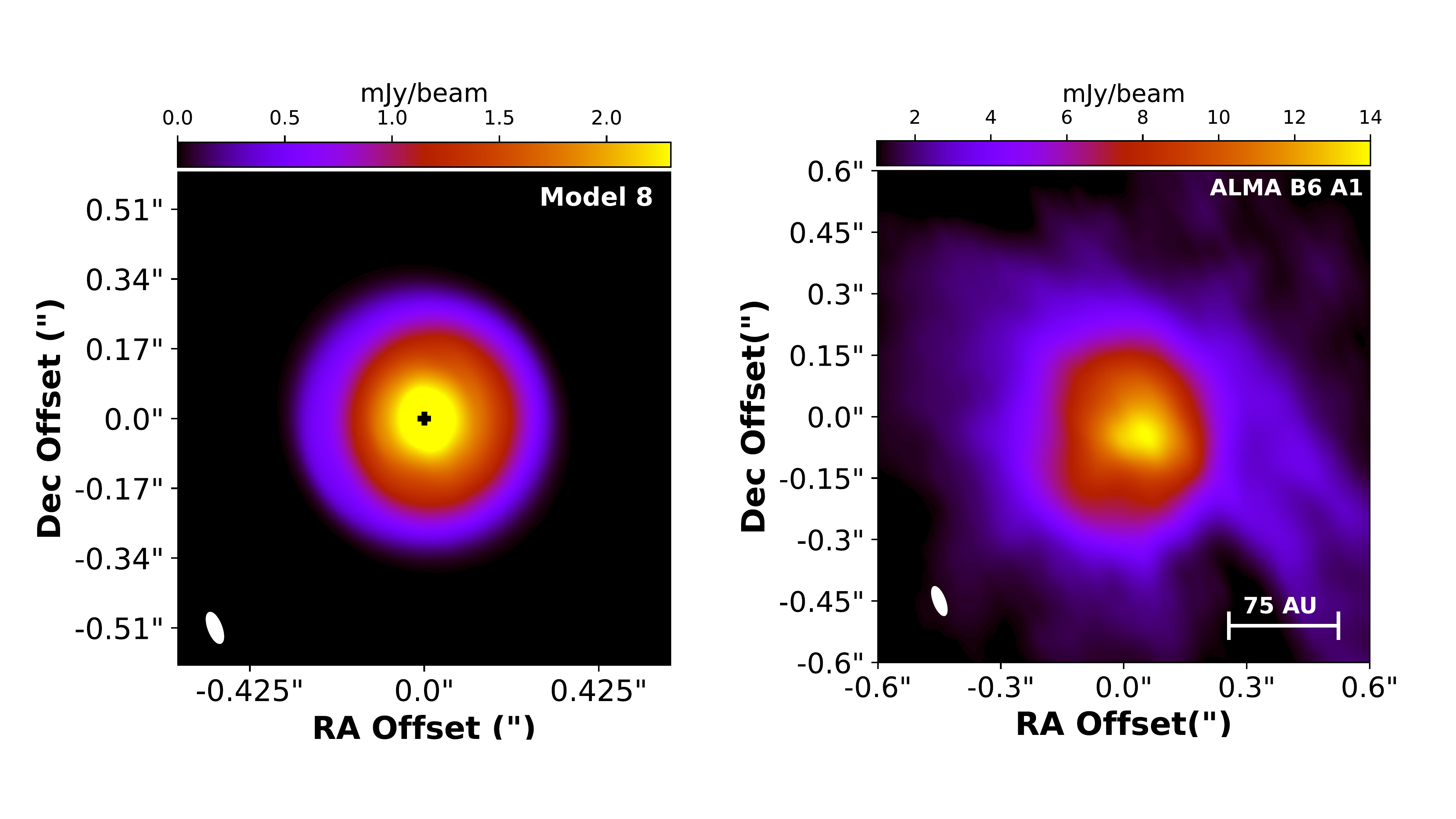}
\caption{ Left: RADMC-3D model 8 convolved with a 0.078" x 0.031" beam. Considering both the major and minor axis cuts, the substructures disappear in the major axis at a scale height of 0.3. The gap radial location is set at 40 au with a small width of 4 au. Right: ALMA continuum unconvolved emission image at 1.2 mm zooming into the A1 source. The respective intensities of each images is shown in the upper colorbars.} \label{fig: IRAS4A1 RADMC-3D best fit}
\end{figure*}
\section{Discussion}

The inferred large scale height (H/R > 0.3) in IRAS4A1 has significant implications for planet formation. Despite the fast settling expected during the disk's lifetime, the optical thickness and asymmetry observed at Band 6 indicate the presence of material with varying grain sizes in higher layers of the disk. This result implies that settling is still ongoing for millimeter-sized particles. Indeed, this state is expected considering the settling timescales (<1 Myr, \citealt{2004A&A...421.1075D}) and the estimated dynamical age of the outflows in IRAS4A (a few 0.01 Myr \citealt{2020A&A...637A..63T}). Furthermore, settling and radial drift are likely acting together during these early stages of dust evolution and growth in the disk.

The large scale height of the disk may also obscure young substructures, as suggested by models, particularly when combined with very narrow substructures measuring less than 4 au in size. While we are unable to directly resolve substructures (i.e., gaps) in 
the disk, our models suggest that some must be present to explain the observed bumps in the radial profile of IRAS4A1. These small-scale features are challenging to observe directly with current resolution capabilities (no substructures observed in IRAS4A1),  but their presence at these early stages could indicate two possibilities. Firstly, if these substructures are caused by planet-disk interactions, it suggests that planets formed nearly instantaneously after the collapse of the molecular cloud. Furthermore, given sufficient time, substructures within protoplanetary disks are expected to widen, resembling those observed in other systems. This widening occurs as planets grow in size by accreting and carving out material from their surroundings.

We note that substructures can potentially arise from mechanisms other than planet-disk interaction. If this is the case, it introduces an intriguing possibility. In this scenario, the substructures initially form early on and are narrow, as indicated by the narrow gap width in the models. Given settling-induced growth of dust particles and other processes occurring within these narrow substructures, substructure formation through alternative mechanisms may itself trigger planet formation within such gaps. Hence, planet formation could take place very early in the disk's evolution, following an evolutionary path similar to the first scenario.

An additional crucial aspect to consider is the large flaring index observed in IRAS4A1. If the "shoulder" position, which served as the basis for defining the gap position for the models, around the range of 20-40 AU is pointing to a substructure, this substructure could be relatively close to the disk mid-plane. For example, smaller scale heights may occur in the center of the disk, meaning that the closer a substructure forms to the center the easier it would be to detect it in a very flared disk. Consequently, any mechanism responsible for carving out these substructures probably starts in the mid-plane and is unable to reach large-scale heights, as seen in other protoplanetary disks when scattered light observations and sub-mm observations with ALMA are compared.

It is important to note, however, that the combination of large scale heights and a large flaring index could still hide further substructures in the outer radii of the IRAS4A1 disk. On the other hand, if there are no substructures in the flared disk of IRAS4A1, planet formation may then occur only at a later stage when larger particles have already settled in the disk midplane, taking into account the timescales required for settling. 

Recently, similar results including the "shoulders", asymmetries, and large scale heights were found in YSOs in studies by the eDisk survey team \citep{2023ApJ...951....8O}. Regardless of the specific dynamics within the IRAS4A1 disk, it is becoming evident that Class 0 Young Stellar Objects (YSOs) exhibit flared disks with significant scale heights, providing valuable insights into the planet formation process. 

\section{Summary and conclusions}

We have shown high-resolution ALMA images (78 mas) of the IRAS4A binary system in Bands 4 and 6. In summary, the key findings of this paper can be outlined as follows:
\begin{itemize}

\item No substructures were detected in either A1 or A2 at the current resolution. 
\item Analysis of spectral indexes and brightness temperatures indicated that A1 is significantly more optically thick than A2. 
\item A multi-wavelength image analysis was carried out showing the dust parameters in A1. The expected values of the dust parameters inferred high temperatures (>50 K), high surface densities (>10 $gcm^{-2}$), and large dust size particles (>30 $\mu$m) at all radius (< 60 au) in the IRAS4A1 disk. In addition, the analysis showed high optical depth in the inner disk in Band 6 and Band 4. 
\item Radiative transfer models using RADMC-3D have shown that a minimum scale height of H/R > 0.3 is adequate to render the substructures invisible and produce an asymmetry in the disks. Moreover, the models that incorporated a narrow gap around 34-50 au and increased flaring index, provided better matches to the observed intensity profiles, suggesting the presence of potential hidden substructures within a very flared disk in the IRAS4A1 system, even in these early stages of disk formation.
\end{itemize}
Observations with high resolution and sensitivity at cm wavelengths with the ngVLA can help unveil any substructure that might exist in IRAS4A1.

\begin{acknowledgements}
We thank the referee for the very constructive comments. We also thank Dominique M. Segura-Cox for the useful discussion. We acknowledge assistance from Allegro, the European ALMA Regional Centre node in the Netherlands. This paper makes use of the following ALMA data: ADS/JAO.ALMA$\#$2018.1.00510.S. ALMA is a partnership of ESO (representing its member states), NSF (USA) and NINS (Japan), together with NRC (Canada), MOST and ASIAA (Taiwan), and KASI (Republic of Korea), in cooperation with the Republic of Chile. The Joint ALMA Observatory is operated by ESO, AUI/NRAO and NAOJ.

EGC acknowledges support from the National Science Foundation through the NSF MPS-Ascend Fellowship Grant number 2213275.
L.W.L. acknowledges support from NSF AST-1910364 and NSF AST-2108794.
\end{acknowledgements}

%
%

%


\bibliographystyle{aa}
\bibliography{bibliography}
\onecolumn
\begin{appendix}

\section{Faint emission between A1 and A2.}
\begin{figure*}[h!]
\centering

\includegraphics[trim=0cm 3cm 0 5cm, clip=true,width=1.00\textwidth]{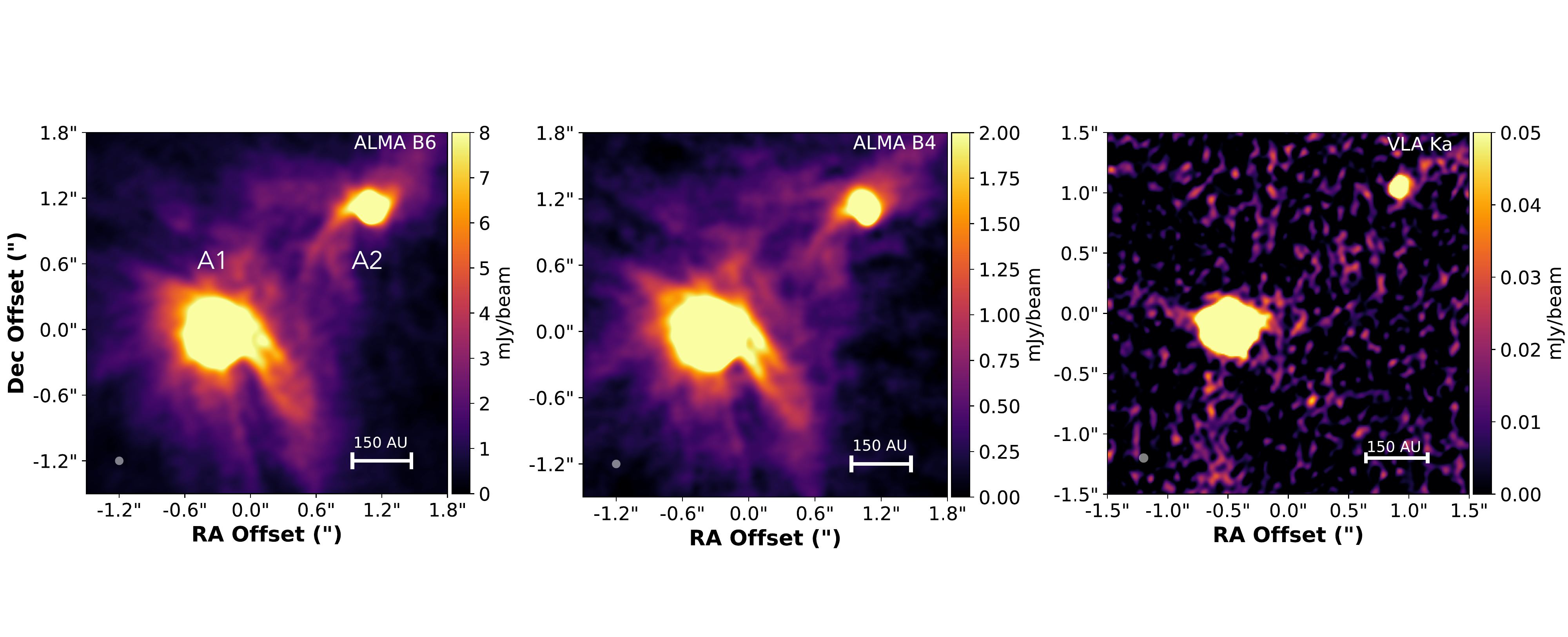}
\caption{Images of IRAS4A at 1.2 mm, 2.1 mm, and 9.1 mm imaged at 78 mas resolution. The color scale was changed to better observed the faint emission between the two targets at B6 and B4.} \label{fig: Continuum Emission convolved at 78 m.a.s FE}
\end{figure*}

\end{appendix}

\end{document}